# Single-Crystal Growth, Structure, and Physical Properties of $Yb_2MoO_6$

Ali Sarikhani[1,†], Santhoshkumar Sundaramoorthy[2], Sadracd Gamez[3], Mathew Pollard[3], Deepak Singh[4], Jeremy Watts[5], Steven Kelley[6], Amitava Choudhury[2,†], Yew San Hor[3,†]

[1] *Materials Research Center, Missouri University of Science and Technology, Rolla, MO, USA*
[2] *Department of Chemistry, Missouri University of Science and Technology, Rolla, MO, USA*
[3] *Department of Physics, Missouri University of Science and Technology, Rolla, MO, USA*
[4] *Department of Physics and Astronomy, University of Missouri, Columbia, MO, USA*
[5] *Department of Materials Science and Engineering, Missouri University of Science and Technology, Rolla, MO, USA*
[6] *Department of Chemistry, University of Missouri, Columbia, MO, USA*

***ORCID:***

***Sarikhani**: 0000-0001-8121-0867*
***Sundaramoorthy**: 0000-0001-5037-0016*
***Pollard**: 0009-0002-2690-4166*
***Singh**: 0000-0003-2700-4063*
***Watts**: 0000-0002-4150-5254*
***Kelley**: 0000-0001-6755-4495*
***Choudhury**: 0000-0001-5496-7346*
***Hor**: 0000-0002-1417-4357*

## Abstract

Rare-earth molybdates are functional oxides of considerable interest because their structural and chemical flexibility enables diverse magnetic, electronic, and optical behavior. Here, millimeter-sized single crystals of $Yb_2MoO_6$ are synthesized, their crystal structure determined and physical properties investigated. Single-crystal X-ray diffraction confirms the formation of $Yb_2MoO_6$ and yielded lattice parameters of $a$ =16.228(2) Å, $b$ = 10.901(3) Å, and $c$ = 5.286(2) Å, in good agreement with previously reported neutron powder diffraction data. Scanning electron microscopy and energy-dispersive spectroscopy show a Yb:Mo ratio close to the expected 2:1 stoichiometry. Optical measurements based on diffuse-reflectance spectra and Tauc analysis yield an estimated optical gap of ~3.28 eV under the direct-allowed transition assumption. Magnetic susceptibility measurements reveal Curie-Weiss-like paramagnetic behavior, with an effective magnetic moment of $\mu_{eff} \approx 4.2\mu_B$ per Yb and an apparent Weiss temperature of $\theta \approx -65.7$ K, consistent with possible antiferromagnetic interactions or correlations, with no clear magnetic-ordering transition observed down to 10 K. These results establish the growth of $Yb_2MoO_6$ single crystals and provide initial insight into their structural, optical, and magnetic properties.

† Corresponding authors: Sarikhani (as5kw@mst.edu), Hor (yhor@mst.edu), Choudhury (choudhurya@mst.edu).

## Introduction

Rare-earth molybdates constitute a chemically and structurally diverse family of oxide materials that have attracted considerable interest because of their wide range of structural, optical, electronic, magnetic, dielectric, and ionic-transport properties [1-4]. An important feature of these compounds is the ability of molybdenum to adopt multiple oxidation states and coordination environments, allowing a large number of compositions and structural arrangements to be stabilized [1,4]. Both higher molybdates containing predominantly $Mo^{6+}$ and reduced compounds containing lower Mo oxidation states have been investigated, with their phase formation and properties depending strongly on composition, oxygen content, and synthesis conditions [1,3,4]. Rare-earth molybdates consequently encompass a variety of structural families, including fluorite-derived, scheelite-related, pyrochlore, and more complex oxide structures [1-4].

The relationship between crystal chemistry and physical behavior is particularly evident in the reduced $Mo^{4+}$-based pyrochlores $R_2Mo_2O_7$. These compounds display a broad range of magnetic and electronic ground states that depend strongly on the identity and ionic radius of the rare-earth ion [5-9]. Early studies established substantial variation in magnetic ordering and electrical transport across the $R_2Mo_2O_7$ series [5,6]. Subsequent work demonstrated that changes in rare-earth ionic radius, and the associated changes in lattice geometry, are closely correlated with the transition between ferromagnetic metallic and insulating or spin-glass states [7,8]. More generally, the pyrochlore lattice provides a well-known framework for geometrically frustrated magnetism because both metal-site sublattices form networks of corner-sharing tetrahedra [9]. These observations emphasize the strong coupling among crystal structure, electronic transport, and magnetic interactions in rare-earth molybdate systems.

Yb-containing molybdates are particularly interesting because $Yb^{3+}$, with its localized $4f^{13}$ electronic configuration, can provide a substantial magnetic moment. In the reduced pyrochlore $Yb_2Mo_2O_7$, previous work has reported semiconducting behavior, significant lattice disorder, magnetic irreversibility, and predominantly antiferromagnetic interactions within the Mo sublattice [10]. $Yb_2MoO_6$, however, is chemically distinct from $Yb_2Mo_2O_7$. Charge balance in $Yb_2MoO_6$ corresponds to $Yb^{3+}$ and $Mo^{6+}$, giving Mo a $4d^0$ electronic configuration. Consequently, unlike $Mo^{4+}$-based pyrochlores in which the Mo sublattice itself contributes localized d-electron magnetism, the localized magnetic response of $Yb_2MoO_6$ is expected to arise primarily from the $Yb^{3+}$ ions. $Mo^{6+}$-containing rare-earth molybdates are also of interest for their optical properties; for example, diffuse-reflectance and Kubelka-Munk analyses have been used to investigate the optical response and bandgap of the related compound $La_2MoO_6$ [11]. $Yb_2MoO_6$ has previously been synthesized and structurally characterized in polycrystalline form.

Alonso et al. prepared a series of $R_2MoO_6$ compounds, including $Yb_2MoO_6$, and established their room-temperature monoclinic C2/c structures using high-resolution neutron powder diffraction [12]. For $Yb_2MoO_6$, the structure consists of discrete $MoO_5$ coordination units incorporated within a rare-earth-oxygen framework [12]. Magnetic properties have also been investigated in related $Mo^{6+}$-containing rare-earth molybdates, such as single-crystalline $KYb(MoO_4)_2$, which exhibits Curie-Weiss-like behavior [13]. Related oxygen-deficient $R_2MoO_{6-\delta}$ compounds, including Yb-containing compositions, have been reported to adopt cubic fluorite-type structures under reducing conditions, highlighting the influence of oxygen stoichiometry and synthesis conditions on phase formation [14]. Additionally, recent investigations of $Gd_2MoO_6$ using first-principles calculations have predicted an indirect electronic bandgap of approximately 1.92 eV, providing further insight into the electronic properties of $Mo^{6+}$-containing rare-earth molybdates [15]. However, to the best of our knowledge, the growth and physical-property characterization of $Yb_2MoO_6$ single crystals have not been reported. The availability of single crystals provides several advantages beyond the earlier powder-based structural study. In particular, it permits direct crystallographic refinement from an individual crystal without orientational averaging inherent to powder diffraction, minimizes complications associated with grain boundaries and polycrystalline microstructure during physical-property measurements, and provides a platform for future direction-dependent studies of anisotropic properties. In the present work, we report the growth of millimeter-sized $Yb_2MoO_6$ single crystals and investigate their crystal structure using single-crystal and powder X-ray diffraction, their microstructure and elemental composition by SEM/EDS, their optical response and bandgap by diffuse-reflectance spectroscopy, and their temperature-dependent magnetic properties.

## Experimental methods

*Synthesis of single crystals*

$Yb_2MoO_6$ single crystals were obtained from a precursor mixture initially prepared with the nominal stoichiometry of $Yb_2Mo_2O_7$. For each synthesis, 1.2500 g of $Yb_2O_3$ (99.998%, Alfa Aesar) and 0.8117 g of $MoO_2$ (99.99%, Alfa Aesar) were mixed in a 1:2 molar ratio. The nominal precursor stoichiometry can be represented as

$$Yb_2O_3 + 2\ MoO_2 \rightarrow Yb_2Mo_2O_7.$$

The powders were thoroughly ground, loaded into a quartz ampoule, and vacuum sealed using a methane-oxygen torch under argon flow. The sealed ampoule was annealed at 1100 °C for 24 h.

The resulting precursor was subsequently ground, pressed into a pellet at 10 MPa for 15 min, transferred to an alumina crucible, and heated in air at 1400 °C for 36 h. This second heat treatment yielded colorless $Yb_2MoO_6$ crystals, typically approximately 1-5 mm in size, together with a smaller amount of yellow secondary material. Because $Yb_2MoO_6$ contains one Mo atom per formula unit, whereas the nominal $Yb_2Mo_2O_7$ precursor contains two, formation of $Yb_2MoO_6$ requires redistribution or loss of the excess Mo. One possible stoichiometric pathway is

$$Yb_2Mo_2O_7 + O_2 \rightarrow Yb_2MoO_6 + MoO_3.$$

This reaction provides a simple mass-balance description in which Mo is oxidized from an average oxidation state of +4 in the nominal $Yb_2Mo_2O_7$ precursor to +6. However, because the identity of the yellow material has not been definitively established, this reaction should not be interpreted as a definitive assignment of the secondary material to $MoO_3$. A quantitative yield of isolated $Yb_2MoO_6$ crystals was not determined because the colorless crystalline material and yellow secondary material were intermixed in the final product. Complete separation and weighing of the two components would have required extensive manual sorting and could not be performed reliably because of the fragility of the crystals. The yellow secondary material was observed primarily on the surfaces of the crystalline pieces, making its complete mechanical separation from the main phase particularly difficult. Nevertheless, suitable colorless crystals were readily distinguishable from the yellow material and could be easily hand-picked using tweezers under a microscope from the bulk crystalline product for subsequent characterization. Qualitatively, the bulk of the product consisted of colorless crystalline material, with a smaller amount of yellow material present. Structural characterization of the isolated colorless crystals by single-crystal and powder X-ray diffraction established that the crystalline product was $Yb_2MoO_6$ rather than the initially targeted $Yb_2Mo_2O_7$.

The synthesis procedure was repeated on three separate occasions, primarily in an effort to obtain additional and larger crystals. $Yb_2MoO_6$ crystals were successfully obtained in each synthesis without modification of the reported procedure, supporting the reproducibility of the crystal-growth method under these experimental conditions. Although suitable $Yb_2MoO_6$ crystals were obtained in all three syntheses, batch-specific counts of crystals suitable for characterization were not recorded. A detailed description of the original synthesis procedure and experimental setup, including illustrations of the experimental steps, is provided in Ref. [16].

*Structural characterization*

Single-crystal X-ray diffraction (SCXRD) was used to characterize the crystal structure of $Yb_2MoO_6$. A small broken piece from a large millimeter-sized crystal was selected from the synthesized batch and mounted on the glass fiber using epoxy resin for diffraction measurements. The single-crystal data sets were collected on Bruker Smart Apex II diffractometer with monochromated Mo Kα radiation (λ= 0.7107 Å) using SMART [17] software at room temperature. The cell refinement and data reduction were carried out using SAINT [18], while the program SADABS [18] was employed to perform the absorption corrections. The crystal structure was solved by direct method using SHELX-97 and different Fourier syntheses [19]. Full matrix least-square refinements against $F^2$ were carried out by SHELXTL-PLUS suite of programs [19]. The atomic positions of all heavy atoms (Yb and Mo) and oxygen were located from the Fourier map and refined isotropically first. Finally, anisotropic refinement on all atoms was carried out, which converged very well with the weighted residual factor, $wR_2$ to a minimum value with no significant residual electron densities remaining. The final refinement was performed with SHELX-2018 incorporated in SHELXle [20]. The final CIF file containing the structure factors has been submitted to CCDC (CSD 2594197) and can be obtained free of charge from The Cambridge Crystallographic Data Centre via https://www.ccdc.cam.ac.uk/structures/.

*Powder X-ray diffraction*

Powder X-ray diffraction (PXRD) data were collected from finely ground, hand-picked colorless crystals that were manually isolated from the crystal-growth product. This data was obtained using PANalytical X'Pert Pro diffractometer using Cu Kα radiation over a 2θ range of 5° to 90° with an average scan speed of 0.094° per second.

*Microstructure and composition*

The microstructure and elemental composition of $Yb_2MoO_6$ were examined using scanning electron microscopy (SEM; TESCAN VEGA3 XMH) equipped with energy-dispersive X-ray spectroscopy (EDS; Bruker XFlash 5100 silicon-drift detector). SEM imaging was performed at an accelerating voltage of 20 kV and a working distance of approximately 16 mm using a secondary-electron detector. EDS measurements were obtained from two separate $Yb_2MoO_6$ crystals, with one region analyzed on each crystal. Elemental EDS mapping was performed to evaluate the spatial distribution of Yb, Mo, and O, and quantitative EDS analysis was used to estimate the metal composition and compare the measured Yb:Mo ratio with the nominal $Yb_2MoO_6$ stoichiometry.

Because oxygen is less reliably quantified by EDS than heavier elements, the compositional analysis was evaluated primarily on the basis of the metal content ratio. For comparison with the ideal $Yb_2MoO_6$ composition, the measured Yb and Mo atomic percentages were normalized to the metal sublattice. The theoretical metal composition of $Yb_2MoO_6$ is 66.7 at% Yb and 33.3 at% Mo, corresponding to a Yb:Mo ratio of 2:1.

*Optical property measurement*

Diffuse-reflectance measurements were carried out using a Cary 5000 UV-Vis-NIR spectrophotometer equipped with a Praying Mantis diffuse-reflectance accessory. A small amount of coarsely ground powder prepared from hand-picked $Yb_2MoO_6$ crystals was used for the measurements, with $BaSO_4$ powder used as the reflectance reference. The optical response was measured over the visible-to-ultraviolet range, and the diffuse-reflectance data were subsequently used for optical-bandgap analysis. The measured optical data were converted from wavelength to photon energy using $h\nu = 1240/\lambda$ where $h\nu$ is the photon energy in eV, and $\lambda$ is the wavelength in nm.

Because the bandgap was estimated using reflectance-based analysis, the absorption coefficient was approximated using the Kubelka-Munk function, which reads as

$$\alpha_{KM} = \frac{(1-R)^2}{2R} \quad [\mathrm{a.u.}] \qquad (1),$$

where $R$ is the reflectance expressed as a fractional value. The optical bandgap was then determined using the Tauc relation (Eq. 2) for a direct allowed transition,

$$(\alpha h\nu)^2 = A\left(h\nu - E_g\right) \quad [eV^2] \qquad (2),$$

where $\alpha$ is the absorption coefficient, $A$ is a proportionality constant, and $E_g$ is the optical bandgap. In the present analysis, $\alpha_{KM}$ was used in place of $\alpha$, and the bandgap was obtained from a plot of $(\alpha_{KM}h\nu)^2$ versus $h\nu$. The linear region near the absorption edge was fitted, and the fitted line was extrapolated to $(\alpha_{KM}h\nu)^2 = 0$ to estimate the optical bandgap [21,22]. For comparison, an indirect-allowed Tauc representation, in which the exponent of $(\alpha h\nu)$ from Eq. 2 is 0.5 rather than 2, was also examined.

*Magnetic property measurement*

The magnetic properties of $Yb_2MoO_6$ single crystals were measured using a Quantum Design Physical Property Measurement System (PPMS). DC magnetization measurements were carried out as a function of temperature under an applied magnetic field. For temperature-dependent susceptibility measurements, magnetization was measured under an applied field of approximately 0.5 T while cooling from 300 to 10 K, and the magnetic susceptibility, $\chi$, was calculated from the recorded magnetization data. For the magnetic measurements, multiple $Yb_2MoO_6$ single crystals were enclosed in plastic film and positioned within a straw sample holder. The individual crystals were randomly oriented relative to the applied magnetic field; therefore, the susceptibility data reported here are not crystallographic-direction resolved. The measured magnetization was converted to molar susceptibility on a per-mole-of-Yb basis using the amount of Yb present in the measured $Yb_2MoO_6$ sample. Because each $Yb_2MoO_6$ formula unit contains two Yb ions, the number of moles of Yb is twice the number of moles of $Yb_2MoO_6$. The effective magnetic moment per Yb was calculated from the fitted Curie constant. To analyze the paramagnetic behavior, the inverse susceptibility was fitted using the Curie-Weiss law, represented in Eq. 3,

$$\chi^{-1} = \frac{1}{C} T - \frac{\theta}{C} \qquad [emu^{-1}\, mol\, Oe] \qquad (3),$$

where $C$ is the Curie constant and $\theta$ is the Weiss temperature. The effective magnetic moment, $\mu_{\mathrm{eff}}$, was calculated from the Curie constant using $\mu_{\mathrm{eff}} = 2.828\sqrt{C}$, where $C$ is expressed in $emu\, K\, mol^{-1} Oe^{-1}$.

## Results and Discussion

### *Single-crystal structure*

The crystal structure of $Yb_2MoO_6$ was examined by single-crystal X-ray diffraction, and the resulting structural features are shown in Figure 1. The asymmetric unit contains three Yb, one Mo and six oxygen sites as shown in Figure 1a. The three Yb atoms adopt eight-coordinated distorted cubic $YbO_8$ coordination with Yb1 – O bond lengths ranging between 2.178(5) and 2.459(5) Å with average $d$(Yb1 – O)$_{av}$ distance of 2.342(2)Å; Yb2 – O distances range between 2.140(5) and 2.589 Å with average $d$(Yb2 – O)$_{av}$ distance of 2.361(2) Å, and $Yb3O_8$ display Yb3 – O distances in the range 2.189(5) and 2.769(5) Å average $d$(Yb3 – O)$_{av}$ distance of 2.385(2) Å (see Table 2). These Yb – O distances corroborate very well with reported data [12]. Mo on the other hand forms a very distorted trigonal pyramidal coordination with Mo – O distances in the range 1.778(5) to 2.222(5), besides there are three other long Mo – O interactions between 2.929(2) and 3.045(2)Å, which are outside the bonding environment. The MoO5 and YbO8 polyhedra are further edge-shared to form the 3-dimensional structure (Figure 1b). Figure 1(c) presents the unit-cell view of $Yb_2MoO_6$, showing the overall three-dimensional framework of the structure and the connectivity among Yb, Mo, and O atoms. The structure can be viewed in terms of Mo-O and Yb-O coordination polyhedra linked through shared oxygen atoms, producing an extended oxide network throughout the crystal. The structural arrangement is further illustrated through representative projections along the $a$-$c$ and $a$-$b$ planes in Figure 1(d) and Figure 1(e), respectively. These projections help visualize the crystallographic organization more clearly than the full unit-cell view alone and emphasize the periodic arrangement of the Yb-Mo-O framework along different directions. The combination of the full structural view and the two selected projections provides a concise but sufficient representation of the crystal structure.

The refined cell parameters and final refinement details of $Yb_2MoO_6$ are listed in Table 1 and the selected bond lengths are provided in Table 2. The resulting structural parameters corroborate very well with previously reported structural parameters refined from neutron diffraction data of $Yb_2MoO_6$ [12].

**Table 1**. Crystal data and refinement parameters for compound $Yb_2MoO_6$.

| **Compound** | **$Yb_2MoO_6$** |
|---|---|
| Empirical formula | Mo $O_6$ $Yb_2$ |
| Formula weight | 538.02 g/mol |
| Temperature | 296(2) K |
| Wavelength | 0.71073 Å |
| Crystal system | Monoclinic |
| Space group | *C2/c* |
| *a* | 16.228(2) Å |
| *b* | 10.901(3) Å |
| *c* | 5.286(2) Å |
| *α* | 90° |
| *β* | 108.553(2)° |
| *γ* | 90° |
| Volume | 886.6(2) $Å^3$ |
| Z | 8 |
| Density (calculated) | 8.061 $Mg/m^3$ |
| Absorption coefficient | 44.571 $mm^{-1}$ |
| F(000) | 1840 |
| Reflections collected | 3925 |
| Independent reflections | 1064 [$R_{(int)}$ = 0.0404] |
| Goodness-of-fit on $F^2$ | 1.109 |
| Final R indices [I>2σ(I)] | $R_1$ = 0.0269, $wR_2$ = 0.0724 |
| R indices (all data) | $R_1$ = 0.0277, $wR_2$ = 0.0731 |
| Largest diff. peak and hole | 2.836 and -2.392 e.$Å^{-3}$ |

**Table 2.** Interatomic distances for the coordination polyhedra of Yb and Mo.

| **Atom – Atom** | **Distance (Å)** | **Atom – Atom** | **Distance (Å)** |
|---|---|---|---|
| Yb1 – O5 | 2.178(5) | Yb3 – O6 | 2.189(5) |
| Yb1 – O5$^{\#1}$ | 2.178(5) | Yb3 – O6$^{\#11}$ | 2.189(5) |
| Yb1 – O4 | 2.354(5) | Yb3 – O5#1 | 2.254(5) |
| Yb1 – O4#1 | 2.354(5) | Yb3 – O5#12 | 2.254(5) |
| Yb1 – O2#2 | 2.377(5) | Yb3 – O6#13 | 2.327(5) |
| Yb1 – O2#3 | 2.377(5) | Yb3 – O6#10 | 2.327(5) |
| Yb1 – O4#4 | 2.459(5) | Yb3 – O2#5 | 2.769(5) |
| Yb1 – O4#5 | 2.459(5) | Yb3 – O2#2 | 2.769(5) |
| Yb2 – O6 | 2.140(5) | Mo1 – O4 | 1.778(5) |
| Yb2 – O5 | 2.230(5) | Mo1 – O3 | 1.787(5) |
| Yb2 – O6#6 | 2.234(5) | Mo1 – O2 | 1.797(5) |
| Yb2 - O1#7 | 2.370(5) | Mo1 – O1 | 1.820(5) |
| Yb2 – O1#4 | 2.386(5) | Mo1 – O5#4 | 2.222(5) |
| Yb2 – O3#8 | 2.444(5) | | |
| Yb2 – O3#9 | 2.493(5) | | |
| Yb2 – O2 | 2.589(5) | | |

#1 -x + 1, y, -z + ½; #2 -x + 1, -y + 1, -z + 1; #3 x, -y + 1, z - ½; #4 -x + 1, -y + 1, -z; #5 x, -y + 1, z + ½; #6 x, -y + 2, z-1/2; #7 x + 1/2, y + 1/2, z + 1; #8 x + 1/2, y + 1/2, z; #9 x + 1/2, -y + 3/2, z + ½; #10 -x + 1, -y + 2, -z + 1; #11 -x + 1, y, -z + 3/2; #12 x, y, z + 1; #13 x, -y + 2, z + ½.

The close correspondence between the present single-crystal lattice parameters and the earlier polycrystalline values is important for two reasons. First, it verifies that the synthesis route employed here successfully produced $Yb_2MoO_6$ single crystals consistent with the known bulk phase reported by Alonso et al. [12]. Second, it indicates that the single-crystal specimen does not exhibit any significant structural deviation from the previously reported polycrystalline compound, despite the different crystal form and processing history. Therefore, the SCXRD analysis establishes a reliable structural foundation for interpreting the compositional, optical, and magnetic properties discussed in the following sections.

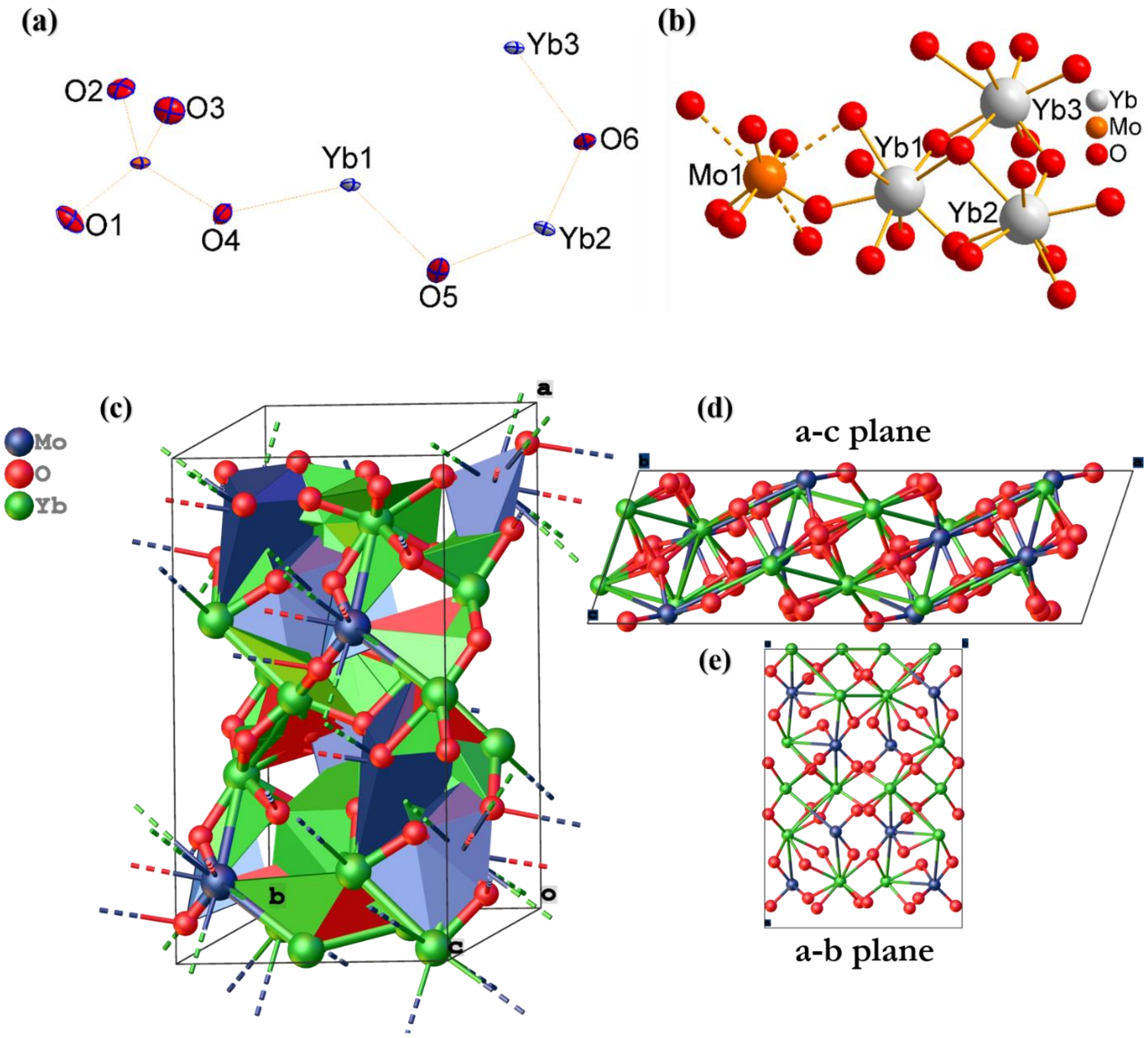

**Figure 1.** Crystal structure of $Yb_2MoO_6$ determined by single-crystal X-ray diffraction: (a) The asymmetric unit showing the thermal ellipsoids of atoms at 60% probability, (b) connectivity between the $MoO_5$ and $YbO_8$ polyhedra, (c) unit-cell view, (d) projection along the $a$-$c$ plane, and (e) projection along the $a$-$b$ plane. Mo, O, and Yb atoms are shown in blue, red, and green, respectively.

*Powder X-ray diffraction*

The experimental diffraction pattern of the isolated colorless crystals matches very well with the pattern simulated using the atomic coordinates obtained from the single-crystal X-ray solution, supporting the phase identity and apparent phase purity of the selected $Yb_2MoO_6$ crystals, as shown in Figure 2. The formation of $Yb_2MoO_6$ from a precursor having the nominal $Yb_2Mo_2O_7$ stoichiometry is consistent with oxidation of Mo during the high-temperature treatment in air. In the nominal $Yb_2Mo_2O_7$ precursor, Mo has an average oxidation state of +4, whereas charge balance in $Yb_2MoO_6$ requires $Mo^{6+}$. Thus, oxidation of Mo during the 1400 °C treatment provides a chemically plausible pathway toward formation of $Yb_2MoO_6$. Because the final $Yb_2MoO_6$ phase contains one Mo atom per formula unit compared with two in the nominal $Yb_2Mo_2O_7$ precursor, formation of $Yb_2MoO_6$ must also involve redistribution or loss of excess Mo. As discussed in the Experimental Methods section, the yellow secondary material observed after the high-temperature treatment may therefore represent a Mo-rich oxide or molybdate phase. However, the identity of this material has not been definitively established. Redistribution and/or volatilization of Mo-containing oxide species during the high-temperature treatment may also contribute to formation of the final $Yb_2MoO_6$ phase. Additional PXRD measurements on a yellow-containing portion of the growth product revealed enhanced diffraction features near 29.9° and 34.4°, suggesting a secondary crystalline contribution (Figure S1, Supplementary Information). Although these features partially overlap with reported reflections of orthorhombic $\alpha$-$MoO_3$, the absence of a clear match with its other characteristic reflections prevents definitive identification of the yellow material.

The major experimental peaks are well reproduced by the pattern simulated from the single-crystal structural model, supporting assignment of the isolated colorless crystals to monoclinic $Yb_2MoO_6$. No major additional reflections attributable to a secondary crystalline phase are evident in the PXRD pattern of the isolated colorless crystals. This observation applies specifically to the hand-picked colorless crystal fraction and does not constitute characterization of the separately observed yellow secondary material. Moreover, the major experimental PXRD peaks (up to 2θ ≈ 60°) and their corresponding calculated Miller indices are summarized in Table 3, showing close agreement between the measured peak positions and those expected for monoclinic $Yb_2MoO_6$.

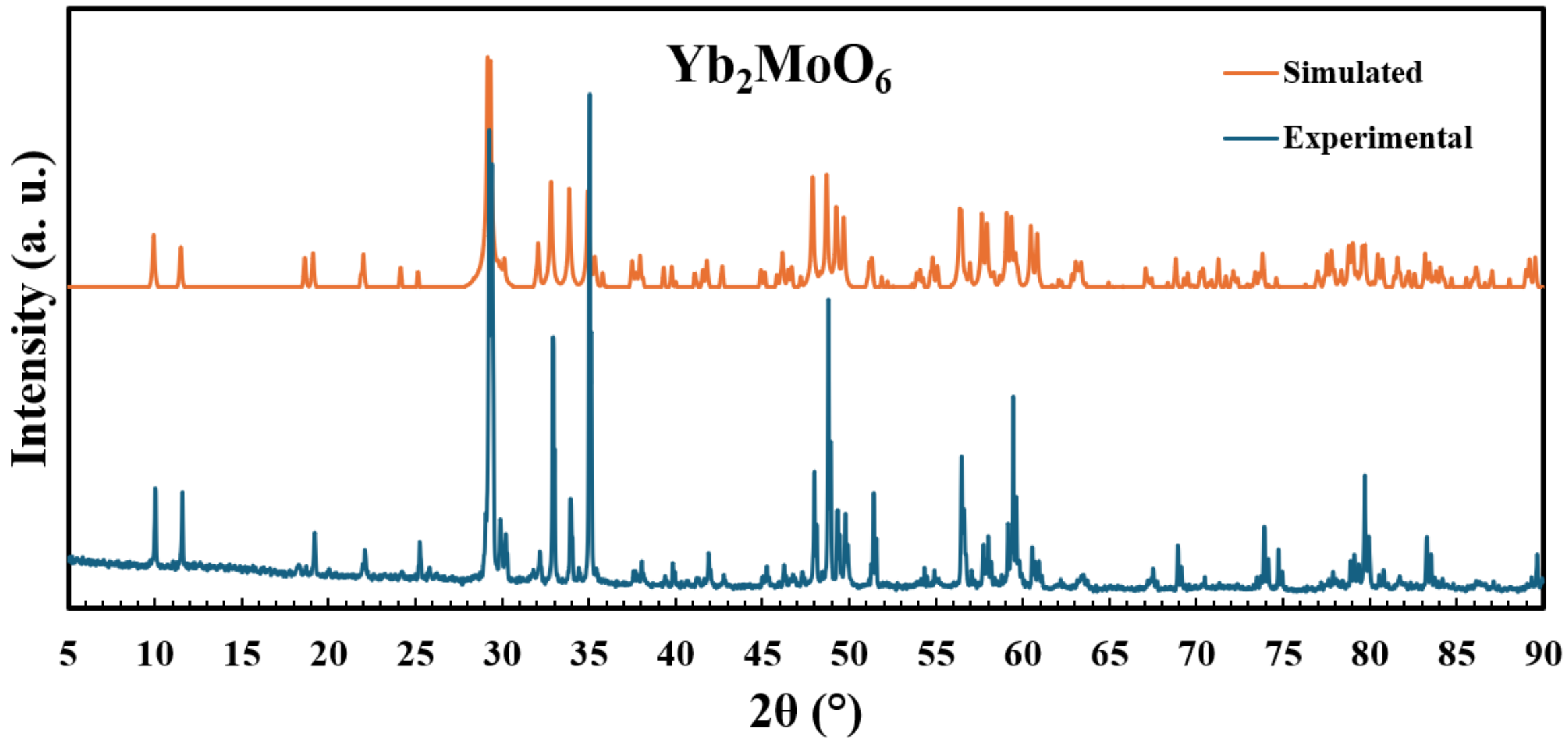

**Figure 2.** Experimental PXRD pattern of ground $Yb_2MoO_6$ crystals compared with the pattern simulated from the single-crystal structural model. The patterns are vertically offset for clarity.

**Table 3**. Selected experimental PXRD peak positions of $Yb_2MoO_6$ and the corresponding calculated reflections and Miller indices based on the monoclinic C2/c structural model.

| Peak position, 2θ (°) | Miller indices, (hkl) |
|---|---|
| **10.05** | (110) |
| **11.60** | (200) |
| **19.23** | (310) |
| **22.13** | $(31\bar{1})$ |
| **29.28** | (221) |
| **29.43** | $(42\bar{1})$ |
| **32.98** | (040) |
| **34.00** | $(20\bar{2})$ |
| **35.10** | (600) |
| **48.03** | $(24\bar{2})$, $(82\bar{1})$ |
| **48.85** | (640) |
| **49.38** | (402) |
| **49.80** | $(80\bar{2})$ |
| **56.53** | (261), (641), $(46\bar{1})$ |
| **57.75** | (023) |
| **58.00** | (113), $(62\bar{3})$ |
| **59.18** | (821), (602) |
| **59.48** | $(13\bar{3})$, $(10\ 2\ \bar{1})$ |
| **60.58** | (442), $(93\bar{2})$ |
| **60.93** | $(84\bar{2})$ |

*Microstructure and composition*

The SEM image of $Yb_2MoO_6$ single crystal, shown in Figure 3, reveals a relatively uniform surface morphology in the analyzed region, without obvious evidence of large secondary-phase particles or severe compositional segregation at the micrometer scale. The corresponding elemental maps further show that Yb, Mo, and O are distributed throughout the scanned area, indicating good overall chemical homogeneity.

Quantitative EDS analysis was used to evaluate the metal composition. For stoichiometric $Yb_2MoO_6$, the theoretical metal-site composition is 66.7 at% Yb and 33.3 at% Mo, corresponding to the expected Yb:Mo ratio of 2:1. In the present work, the average measured metal-site composition was 66.0 ± 1.0 at% Yb and 34.0 ± 1.0 at% Mo, which is close to the ideal stoichiometry. Here, the ± 1.0 at% values represent the reproducibility between the different measured regions rather than the full absolute uncertainty of the EDS technique. The small deviation from the theoretical ratio is within the common error range in EDS quantification, where uncertainties of a few atomic percent are typical depending on counting statistics, calibration, matrix corrections, and local surface conditions.

Although oxygen mapping is useful for confirming its presence and approximate spatial distribution, oxygen quantification by EDS is generally less reliable than that of heavier elements such as Yb and Mo. This is because oxygen emits low-energy characteristic X-rays that are more easily absorbed and are more sensitive to background subtraction, detector response, and surface condition effects. Therefore, the compositional assessment is more appropriately based on the Yb:Mo content ratio rather than the absolute oxygen content. Taken together, the SEM image, elemental maps, and EDS quantification indicate that the synthesized $Yb_2MoO_6$ is chemically homogeneous and close to the intended 2:1 Yb:Mo stoichiometry.

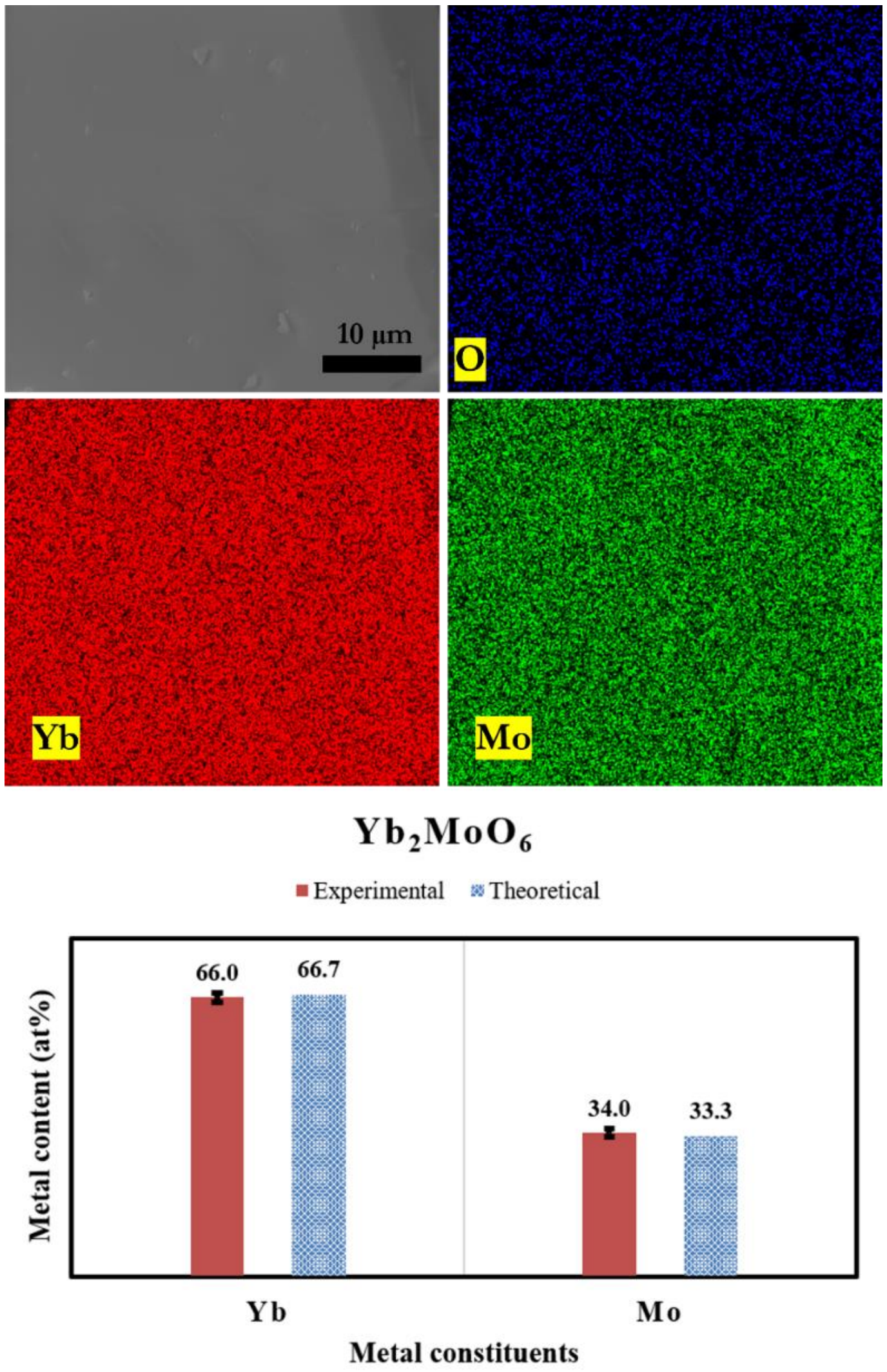

**Figure 3.** SEM image, EDS elemental maps, and metal-site composition of $Yb_2MoO_6$. The elemental maps show a homogeneous distribution of Yb, Mo, and O in the analyzed region. Quantitative EDS analysis gives a metal-site composition close to the theoretical 2:1 Yb:Mo ratio.

*Optical properties*

Figure 4 shows the Kubelka-Munk transformed optical response of $Yb_2MoO_6$ as a function of photon energy in order to obtain the absorption coefficient. At low photon energies, the transformed absorption-related quantity $\alpha_{KM}$ remains close to zero, indicating weak absorption in this region. A weak feature near 1.5 eV may be associated with an intra-4f transition of $Yb^{3+}$. In contrast, a pronounced increase in $\alpha_{KM}$ occurs near ∼3 eV, corresponding to the onset of strong optical absorption. This behavior is consistent with the transparent appearance of $Yb_2MoO_6$ and suggests that the compound is a wide-bandgap oxide.

Above the absorption edge, $\alpha_{KM}$ continues to increase and reaches a broad maximum at higher photon energies, indicating stronger optical transitions well above the threshold region. The weak absorption at low photon energies together with the sharp near-edge increase supports semiconducting or insulating behavior. To quantify the optical bandgap more rigorously, Tauc analysis was subsequently performed using the near-edge region of the spectrum, as discussed in Figure 5.

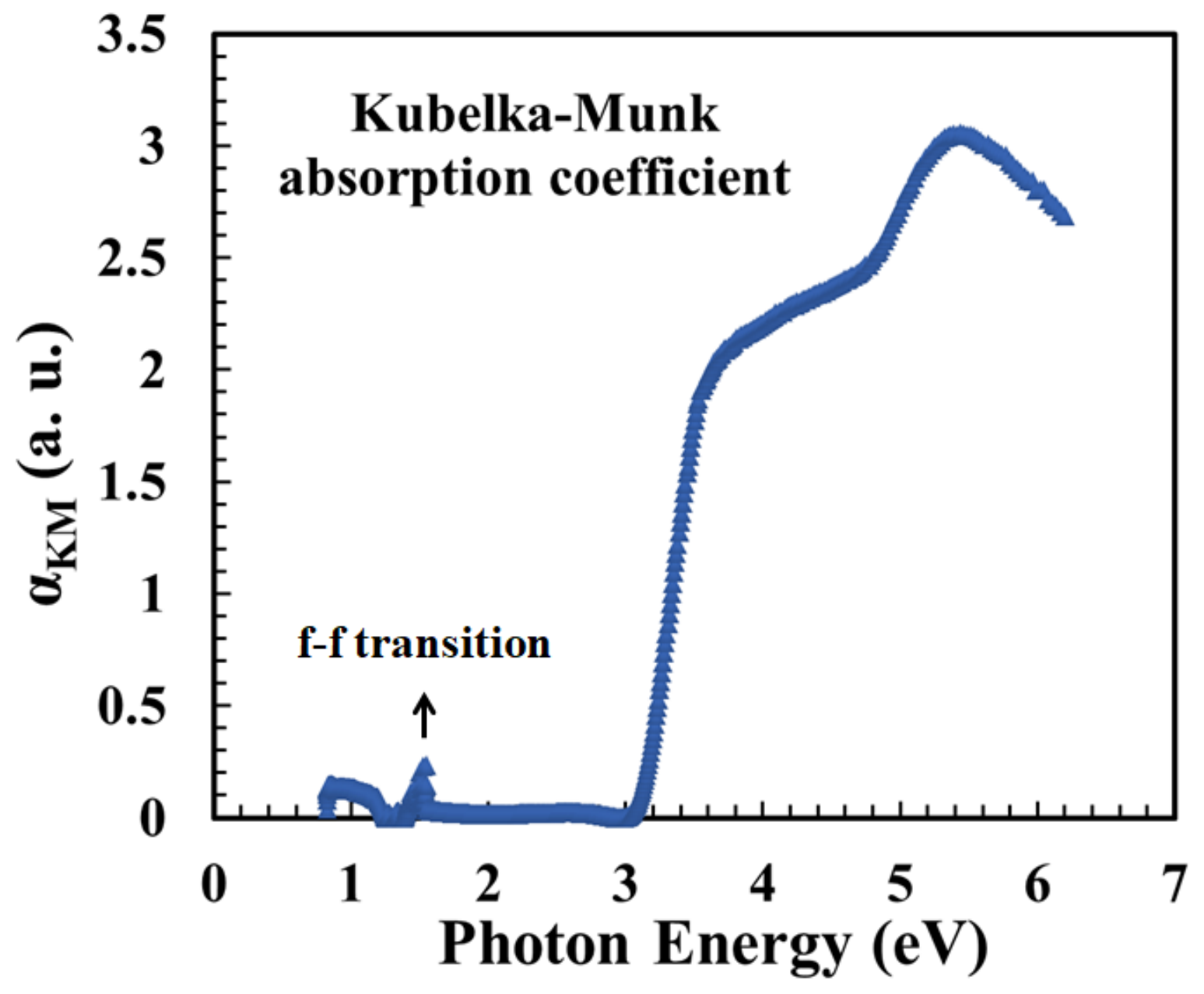

**Figure 4.** Kubelka-Munk transformed optical response of $Yb_2MoO_6$ as a function of photon energy. A weak low-energy feature is observed possibly due to an intra-4f transition of $Yb^{3+}$, while the sharp increase near ~3 eV marks the onset of strong optical absorption.

To quantify the optical bandgap of $Yb_2MoO_6$, the near-edge optical absorption was analyzed using the Tauc method, as shown in Figure 5, and confirmed the presence of large bandgap. The Tauc plot in Figure 5(a) was constructed by plotting $(\alpha h\nu)^2$ as a function of photon energy, assuming a direct allowed transition. A pronounced rise in the Tauc signal is observed above the absorption edge, consistent with the onset of strong interband optical absorption. The broad shape of the full curve also showed that only a limited portion near the absorption threshold is appropriate for bandgap fitting, whereas the higher-energy region deviates from linearity.

Accordingly, a linear fit was applied only to the near-edge region of the Tauc plot, as shown in Figure 5(b). The fitted relation was $(\alpha h\nu)^2 = 166.51\, h\nu - 545.32$, with a coefficient of determination of $R^2 = 0.998$, indicating excellent linearity in the selected fitting range. Extrapolation of this fit to $(\alpha h\nu)^2 = 0$ yielded an optical bandgap of $E_g = 3.28$ eV. This value is consistent with the weak low-energy absorption and the transparent appearance of the synthesized material, and it confirms that $Yb_2MoO_6$ is a wide-bandgap oxide. The direct-allowed linear fit was performed over a photon-energy interval of approximately 3.351-3.503 eV. To evaluate the sensitivity of the estimated optical gap to the fitting interval, additional fits were performed over neighboring intervals, yielding approximately identical optical gaps with deviations of less than 0.01 eV. This indicates that the estimated optical gap is relatively insensitive to modest changes in the fitting interval.

The expanded view in Figure 5(b) highlights the linear near-edge region used for bandgap determination and avoids fitting non-linear portions of the Tauc curve. An indirect-allowed Tauc representation was also examined for comparison; however, it exhibited a pronounced kink in slope near approximately 3 eV rather than a single well-defined linear absorption-edge region. The corresponding indirect-allowed Tauc plot is provided in Figure S2 of the Supplementary Information. The indirect model was therefore considered less suitable for reliable bandgap estimation from the present data. Accordingly, the value of approximately 3.28 eV is regarded as an optical-gap estimate obtained under the direct-allowed transition assumption rather than as definitive evidence of a direct fundamental bandgap.

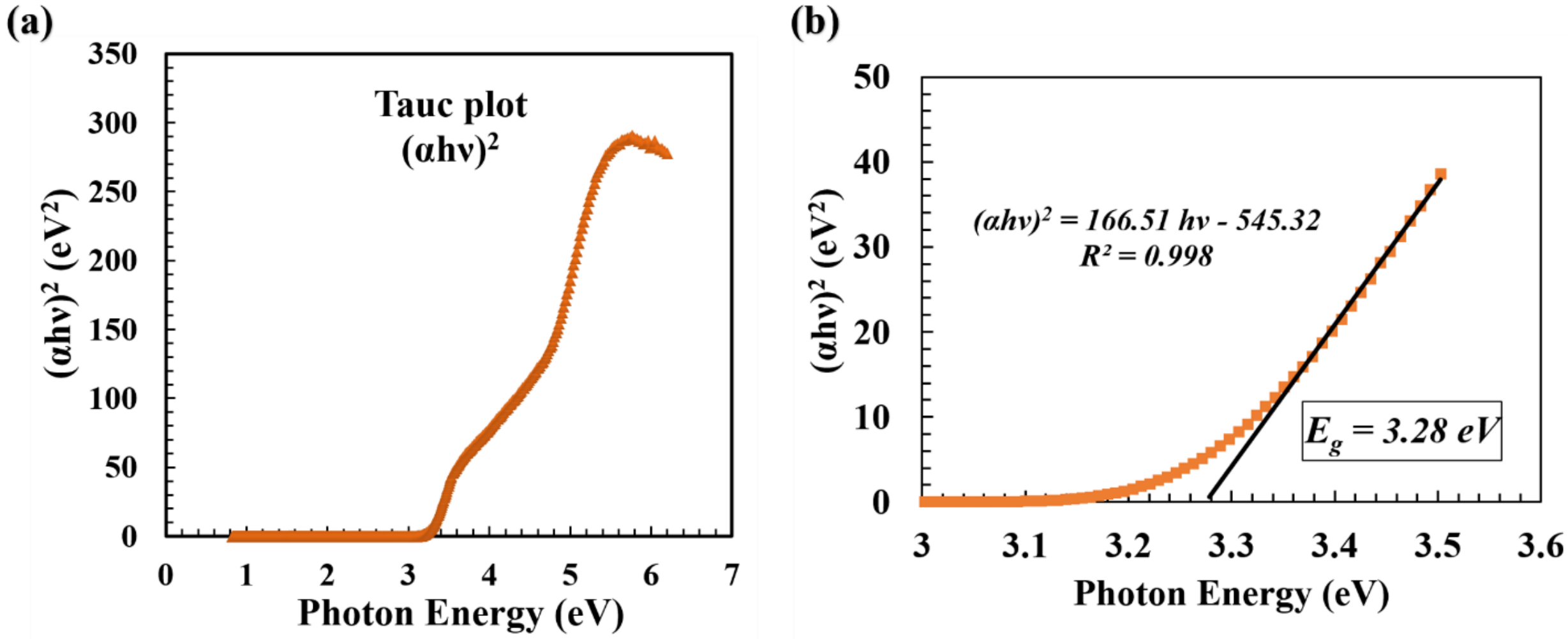

**Figure 5.** Optical bandgap analysis of $Yb_2MoO_6$: (a) Tauc plot constructed from the Kubelka-Munk-transformed optical data and (b) linear fitting of the near-edge region used to determine the optical bandgap. The extrapolated intercept gives $E_g = 3.28$ eV.

*Magnetic properties*

The temperature dependence of the magnetic susceptibility of $Yb_2MoO_6$ measured at $H$ = 0.5 T is shown in Figure 6(a), together with the corresponding derivative $d\chi/dT$, representing a paramagnetic trend. The magnetic susceptibility increases strongly upon cooling, consistent with Curie-Weiss-like paramagnetic behavior. Plotting $d\chi/dT$ is useful for visually emphasizing subtle changes in slope or weak anomalies that may be difficult to identify directly from the $\chi(T)$ curve. In the present data, a weak feature is observed near approximately 50 K. Similar anomalies near this temperature have been reported in low-temperature magnetization measurements and attributed to trace oxygen contamination associated with condensed oxygen in the sample environment [23-25]. Because the origin of the present feature was not independently verified, we consider this an extrinsic possibility rather than assigning the anomaly to an intrinsic magnetic transition of $Yb_2MoO_6$.

To evaluate the intrinsic magnetic response, the inverse susceptibility was analyzed using the Curie-Weiss law, as shown in Figure 6(b). The Curie-Weiss fit was performed over 147.3-257.9 K, where the inverse susceptibility exhibits an approximately linear temperature dependence. At higher temperatures, the inverse susceptibility begins to deviate from linearity, while a more pronounced deviation is also observed below approximately 100 K. The Curie-Weiss fit and the full inverse-susceptibility curve is provided in Figure S3 of the Supplementary Information. Additional fits over neighboring temperature intervals produced only minor variations in the Curie-Weiss parameters, as summarized in Table S1, supporting the relative stability of the extracted parameters within this temperature range. Within the fitting range, the linear fit gave the relation of $\chi^{-1} = 0.462\,T + 30.375$ with a high coefficient of determination of $R^2 = 0.998$. From this fit, the Curie constant was determined to be $C \approx 2.16\ emu\ K\ mol^{-1} Oe^{-1}$ per Yb, corresponding to an effective magnetic moment of $\mu_{\text{eff}} \approx 4.2\mu_B$ per Yb, while the Weiss temperature was estimated as $\theta \approx$ -65.7 K. The effective magnetic moment per Yb shows slightly lower value than the theoretical magnetic moment of $Yb^{3+}$ ($4f^{13}$) using $\mu_{\text{eff}} = g\sqrt{J(J+1)}$. The negative Weiss temperature is consistent with possible antiferromagnetic interactions or correlations among the $Yb^{3+}$ moments. However, crystal-field effects associated with $Yb^{3+}$ may also contribute to the apparent Weiss temperature obtained from the Curie-Weiss fit. Therefore, the magnitude of $\theta$ should not be interpreted solely in terms of magnetic exchange interactions. No clear magnetic-ordering transition is observed down to 10 K. For comparison, single-crystalline $KYb(MoO_4)_2$ exhibits a reported Weiss temperature of approximately -52 K, comparable in sign and magnitude to that obtained for $Yb_2MoO_6$. However, differences in crystal structure and crystal-field environments prevent a direct comparison of the underlying exchange interactions. The present measurements establish the Curie-Weiss-like magnetic response of $Yb_2MoO_6$, providing a basis for future investigations of its low-temperature magnetic behavior.

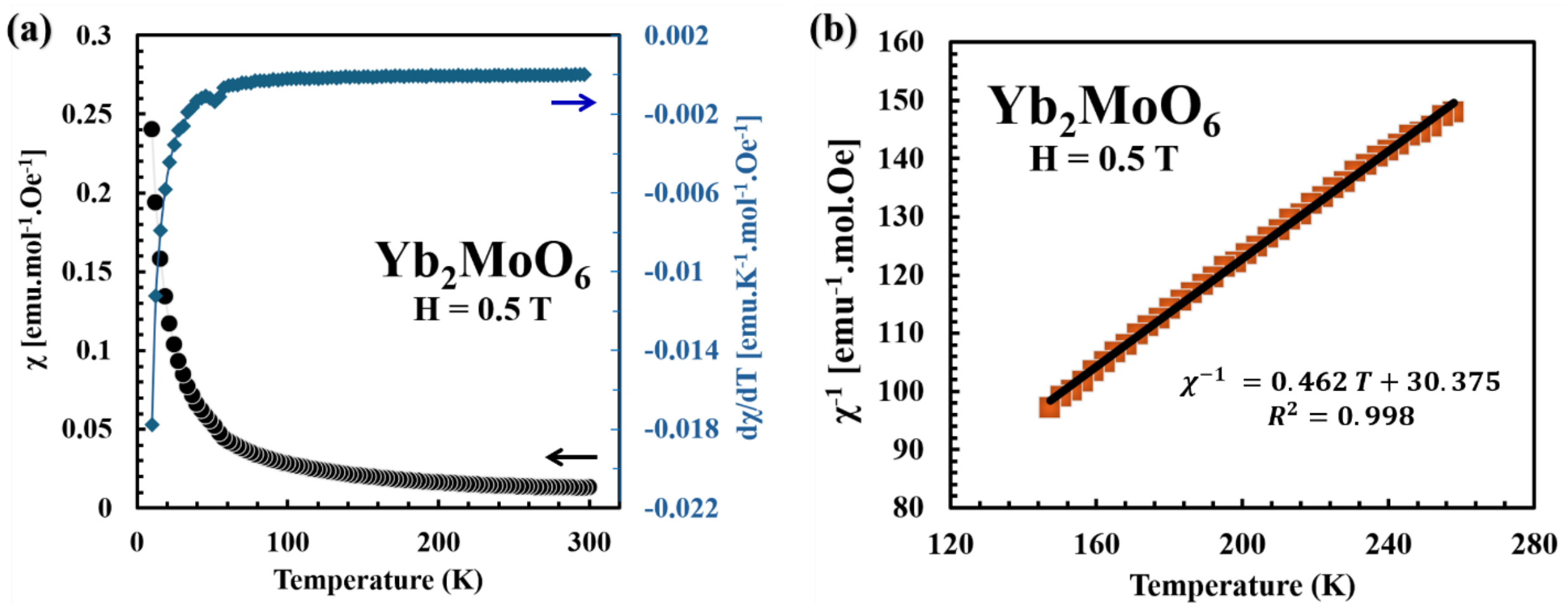

**Figure 6.** Magnetic properties of $Yb_2MoO_6$: (a) temperature-dependent magnetic susceptibility $\chi(T)$ measured at $H$ = 0.5 T together with $d\chi/dT$, and (b) Curie-Weiss analysis of the inverse susceptibility in the paramagnetic region. The linear fit gives $C \approx 2.16\ emu\ K\ mol^{-1} Oe^{-1}$, $\mu_{eff} \approx 4.2\mu_B$ per Yb, and $\theta \approx$ -65.7 K.

## Conclusions

In summary, $Yb_2MoO_6$ single crystals were successfully synthesized and their structural, compositional, optical, and magnetic properties were investigated. Single-crystal X-ray diffraction confirmed formation of the $Yb_2MoO_6$ phase and yielded lattice parameters in close agreement with previously reported polycrystalline data, supporting the structural consistency of the grown crystals. SEM/EDS analysis showed a homogeneous distribution of Yb, Mo, and O, and the measured metal composition was close to the expected 2:1 Yb:Mo stoichiometry.

Optical characterization using Kubelka-Munk and Tauc analysis indicated that $Yb_2MoO_6$ is a wide-bandgap oxide with an estimated optical gap of approximately 3.28 eV under the direct-allowed transition assumption. Magnetic susceptibility measurements showed Curie-Weiss-like paramagnetic behavior, and Curie-Weiss analysis gave an effective magnetic moment of approximately $4.2\mu_B$ per Yb and a negative Weiss temperature of about -65.7 K, consistent with possible antiferromagnetic interactions or correlations. No clear magnetic-ordering transition was observed down to 10 K. Overall, this work establishes $Yb_2MoO_6$ as a single-crystalline rare-earth molybdate with a wide optical bandgap and antiferromagnetically correlated magnetic behavior, providing a foundation for future studies of its intrinsic physical properties.

**Acknowledgements**

The authors are grateful for the support by Materials Research Center at Missouri S&T. We gratefully acknowledge Prof. William Fahrenholtz and Prof. Gregory Hilmas of the Department of Materials Science and Engineering at Missouri S&T for their support. This work was supported in part by the National Science Foundation under Grant No. NSF MRI 2407920.

**Contributions**

**Sarikhani** (Conceptualization, Methodology, Visualization, Validation, Writing - original draft, Writing - review & editing, Data curation); **Sundaramoorthy** (Methodology, Data curation, Writing - review & editing); **Gamez** (Methodology); **Pollard** (Conceptualization, Methodology); **Singh** (Supervision, Writing - review & editing); **Watts** (Methodology, Writing - review & editing); **Kelley** (Methodology, Data curation); **Choudhury** (Conceptualization, Methodology, Visualization, Supervision, Validation, Writing - original draft, Writing - review & editing); **Hor** (Conceptualization, Methodology, Visualization, Supervision, Validation, Writing - review & editing).

**Data Availability**

The CIF file and checkcif-validation have been submitted as supplementary materials and CIF file can also be obtained from the CCDC by quoting CSD number, CSD 2594197. Any other data reported in this manuscript can be obtained on request.

# *Supplementary Material*

*Powder X-ray diffraction*

To further investigate the yellow secondary material observed during crystal growth, an additional PXRD measurement was performed on a portion of the product containing the yellow material. As shown in Figure S1, the diffraction pattern remains predominantly consistent with $Yb_2MoO_6$. However, a pronounced enhancement of the reflection near 29.9° and an additional feature near 34.4° are observed compared with the isolated colorless crystal fraction. These differences suggest the presence of a secondary crystalline contribution, although its identity cannot be definitively established from the present measurements. The additional diffraction features near 29.9° and 34.4° are close to reported reflections of orthorhombic α-$MoO_3$ [1]. However, other characteristic reflections of this phase are not clearly resolved in the measured pattern, preventing its definitive identification. Therefore, the secondary material remains unidentified, although the possible presence of a Mo-rich oxide cannot be excluded.

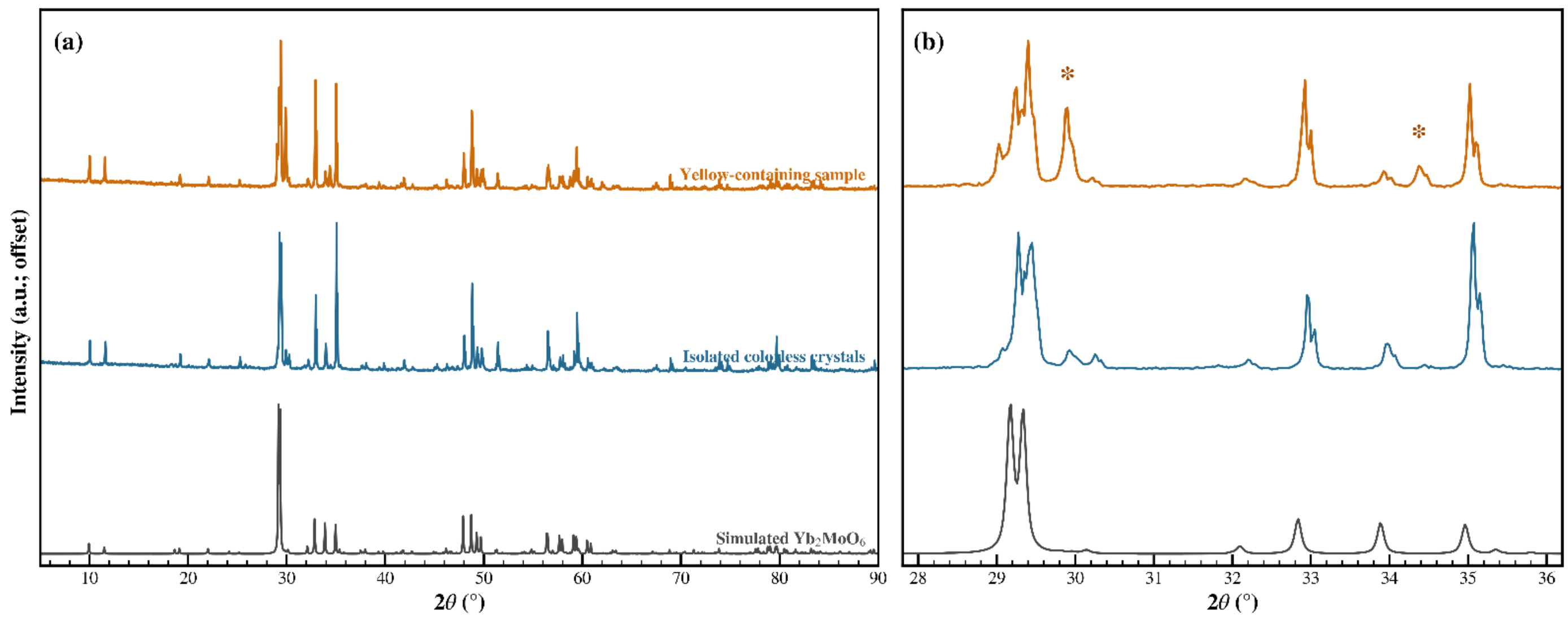

**Figure S1.** Comparison of PXRD patterns of simulated $Yb_2MoO_6$, isolated colorless $Yb_2MoO_6$ crystals, and a yellow-containing portion of the growth product: (a) full diffraction patterns and (b) enlarged view of the 28-36° region. Asterisks indicate reflections exhibiting additional contributions in the yellow-containing sample. The patterns are independently normalized and vertically offset for clarity.

*Optical properties*

For comparison with the direct-allowed Tauc analysis presented in the main text, the indirect-allowed Tauc representation of $Yb_2MoO_6$ is shown in Figure S2. The plot exhibits a pronounced kink in slope near approximately 3 eV rather than a single well-defined linear absorption-edge region, making the selection of a reliable fitting interval less straightforward. The weaker feature at lower photon energies may be associated with intra-4f transitions of $Yb^{3+}$, as discussed in the main text.

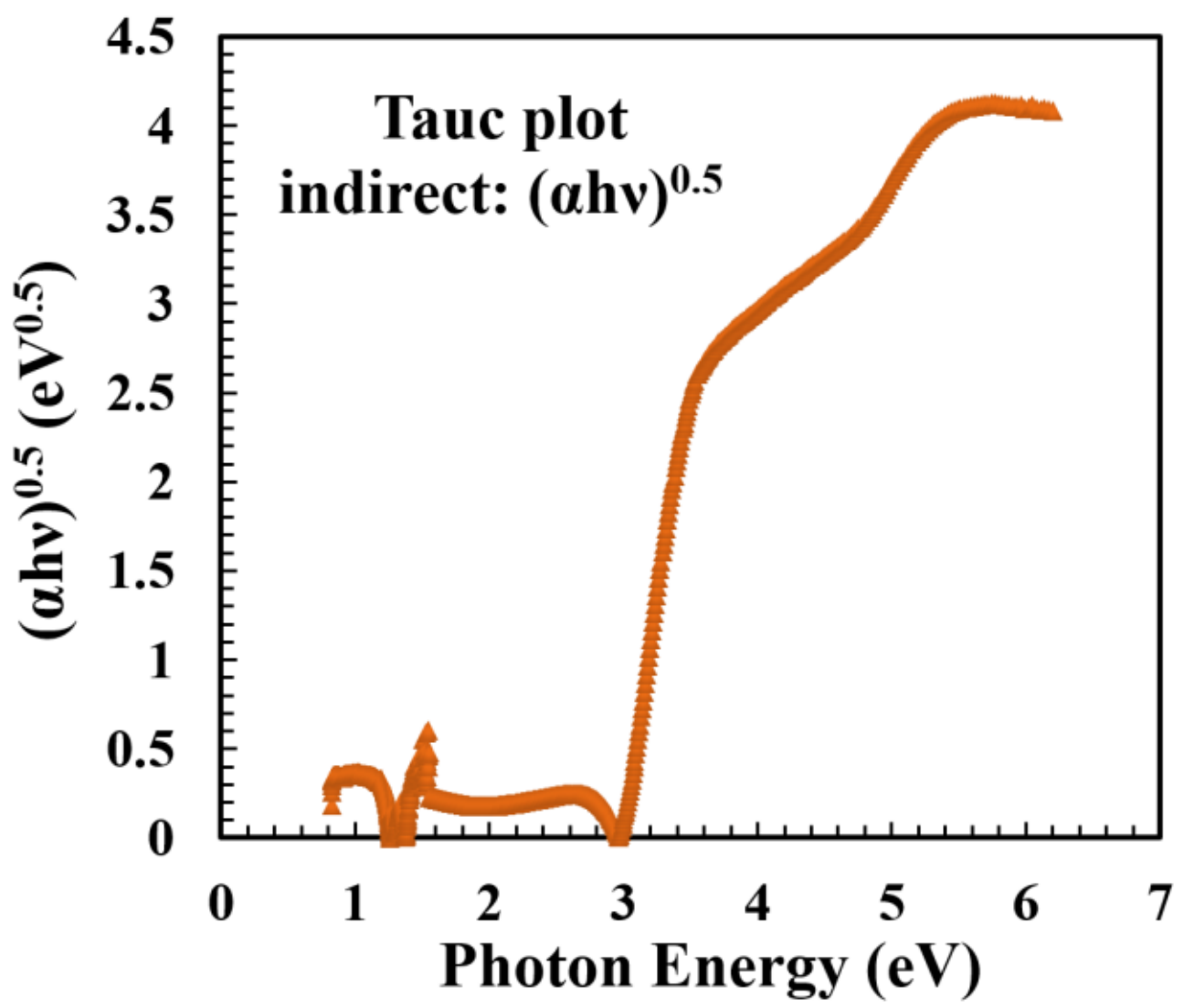

**Figure S2.** Indirect-allowed Tauc plot of $Yb_2MoO_6$ obtained from the Kubelka-Munk-transformed diffuse-reflectance data. The plot exhibits a pronounced change in slope near approximately 3 eV rather than a single well-defined linear absorption-edge region, making reliable determination of the optical gap using the indirect-allowed transition model less straightforward. The weak feature at lower photon energies may be associated with intra-4f transitions of $Yb^{3+}$.

*Magnetic properties*

The full temperature dependence of the inverse magnetic susceptibility of $Yb_2MoO_6$ over 10-300 K is shown in Figure S3, with the original Curie-Weiss fitting interval highlighted. To evaluate the sensitivity of the fitted parameters to the selected temperature range, additional linear fits were performed over neighboring intervals, as summarized in Table S1. The resulting Curie constants range from 2.134 to 2.175 emu K $mol^{-1}$ Yb $Oe^{-1}$, while the Weiss temperatures vary from -61.8 to -67.0 K. These relatively small variations indicate that the extracted Curie-Weiss parameters are not strongly dependent on modest changes in the fitting interval.

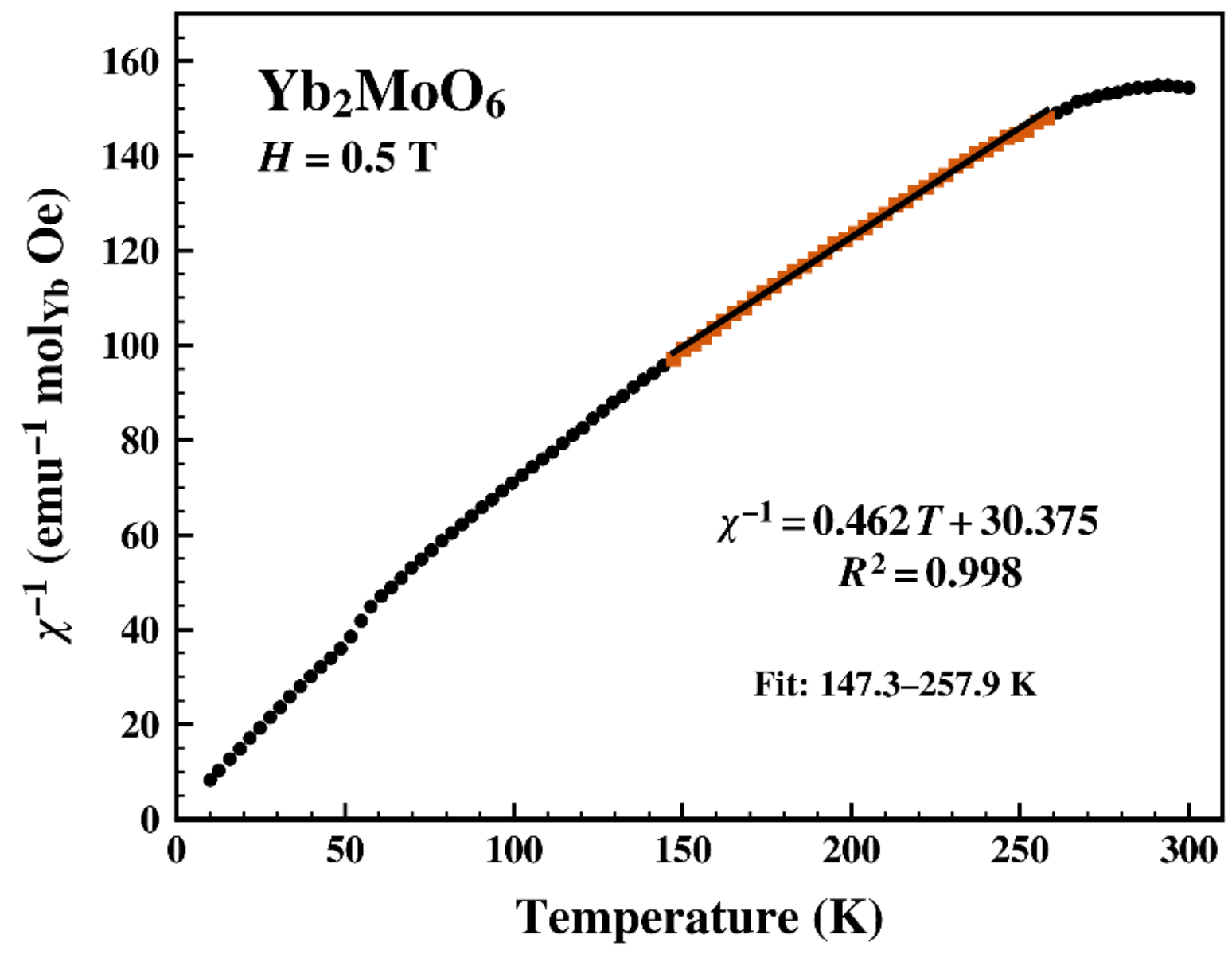

**Figure S3.** Full temperature dependence of the inverse magnetic susceptibility of $Yb_2MoO_6$ measured under an applied field of 0.5 T. The highlighted data points indicate the original Curie-Weiss fitting interval (147.3-257.9 K), and the solid black line represents the corresponding linear fit.

**Table S1**. Dependence of the Curie-Weiss fitting parameters of $Yb_2MoO_6$ on the selected temperature interval. The Curie constant (C) and effective magnetic moment ($\mu_{eff}$) are normalized per Yb ion.

| Fitting interval (K) | C (emu K $mol^{-1}$ $Oe^{-1}$/Yb) | θ (K) | $\mu_{eff}$ ($\mu_B$/Yb) | $R^2$ |
|---|---|---|---|---|
| 147.293-257.905 (original) | 2.163 | -65.71 | 4.160 | 0.99822 |
| 141.332-257.905 | 2.147 | -63.60 | 4.144 | 0.99817 |
| 135.329-257.905 | 2.134 | -61.80 | 4.131 | 0.99818 |
| 150.293-257.905 | 2.172 | -66.91 | 4.168 | 0.99828 |
| 147.293-260.898 | 2.175 | -67.03 | 4.170 | 0.99800 |